\documentclass[conference,a4paper]{IEEEtran}
\IEEEoverridecommandlockouts
\usepackage{cite}
\usepackage{amsmath,amssymb,amsfonts,flushend}
\usepackage{algorithm}
\usepackage{algpseudocode}
\usepackage{graphicx}
\usepackage{textcomp}
\usepackage{xcolor}
\def\BibTeX{{\rm B\kern-.05em{\sc i\kern-.025em b}\kern-.08em
    T\kern-.1667em\lower.7ex\hbox{E}\kern-.125emX}}
\begin{document}

\title{Performance Optimization for Movable Antenna Enhanced MISO-OFDM Systems}
\author{\IEEEauthorblockN{Ruixi Feng\IEEEauthorrefmark{1}, Weidong Mei\IEEEauthorrefmark{2}, Lele Lu\IEEEauthorrefmark{1},  Xin Wei\IEEEauthorrefmark{2}, Zhi Chen\IEEEauthorrefmark{2}, Zhen Gao\IEEEauthorrefmark{3}, and Boyu Ning\IEEEauthorrefmark{2}}  \IEEEauthorblockA{\IEEEauthorrefmark{1}Glasgow College,}  \IEEEauthorblockA{\IEEEauthorrefmark{2}National Key Laboratory of Wireless Communications, \\University of Electronic Science and Technology of China, Chengdu 611731, China.}
\IEEEauthorblockA{\IEEEauthorrefmark{3}State Key Laboratory of CNS/ATM,
Beijing Institute of Technology, Beijing 100081, China.}
Emails: 2022190905006@std.uestc.edu.cn, 
wmei@uestc.edu.cn,
ll.lu@std.uestc.edu.cn, \\
xinwei@std.uestc.edu.cn
chenzhi@uestc.edu.cn, gaozhen16@bit.edu.cn, boydning@outlook.com
}

\maketitle

\begin{abstract}
Movable antenna (MA) technology offers a flexible approach to enhancing wireless channel conditions by adjusting antenna positions within a designated region. While most existing works focus on narrowband MA systems, this paper investigates MA position optimization for an MA-enhanced multiple-input single-output (MISO) orthogonal frequency-division multiplexing (OFDM) system. This problem appears to be particularly challenging due to the frequency-flat nature of MA positioning, which should accommodate the channel conditions across different subcarriers. To overcome this challenge, we discretize the movement region into a multitude of sampling points, thereby converting the continuous position optimization problem into a discrete point selection problem. Although this problem is combinatorial, we develop an efficient partial enumeration algorithm to find the optimal solution using a branch-and-bound framework, where a graph-theoretic method is incorporated to effectively prune suboptimal solutions. In the low signal-to-noise ratio (SNR) regime, a simplified graph-based algorithm is also proposed to obtain the optimal MA positions without the need for enumeration. Simulation results reveal that the proposed algorithm outperforms conventional fixed-position antennas (FPAs), while narrowband-based antenna position optimization can achieve near-optimal performance.
\end{abstract}

\section{Introduction}
Movable antenna (MA) technology has recently attracted significant attention in wireless communications due to its ability to dynamically adjust the positions of transmit and receive antennas within a confined region\cite{zhu2025tutorial,zhu2024opportunity,ningimplement}. Unlike conventional fixed-position antennas (FPAs), MAs can adaptively avoid deep-fading positions, thereby improving wireless channel conditions to enhance data transmission rate\cite{zhu2024performance2,shao6DMA}. Additionally, MAs enable flexible adjustment of antenna geometry, facilitating more precise wireless sensing \cite{masensing,wang2025antenna} and more effective array signal processing \cite{zhu2023movable,wang2024flexible,mamultibeam}.

Motivated by the promising benefits of the MA technology, prior studies have delved into the antenna position optimization under various system setups. Readers can refer to a recent tutorial paper \cite{zhu2025tutorial} for a detailed literature review. It is worth pointing out that the highly nonlinear relationship between channel responses and antenna positions gave rise to a significant challenge in optimizing antenna positions. To tackle this problem, the authors in \cite{huttd,mamimocap,husecure,Wang2024a,zhu2024multi} proposed a variety of gradient-based algorithms (e.g., successive convex approximation and gradient descent) and heuristic algorithms (e.g., particle swarm optimization). However, as these proposed algorithms may be easily trapped at local optimal solutions, some recent works, e.g., \cite{mei2024graph,ma2025robust,wumultidis,mei2024secure,wei2024,weitcom} have proposed a discrete sampling approach, which discretizes the transmit or receive region into a multitude of sampling points. Based on the channel state information (CSI) for each sampling point, the antenna positions can be optimized by solving a point selection problem subject to the constraints on minimum antenna spacing to avoid mutual coupling. For example, in the single- and multi-user multiple-input single-output (MISO) scenarios, the authors in \cite{mei2024graph,ma2025robust} and \cite{wumultidis} have proposed a graph-based algorithm and a generalized Bender’s decomposition method to obtain the optimal discrete MA positions, respectively. 

\begin{figure}[!t]
  \centering
  \includegraphics[width=\linewidth]{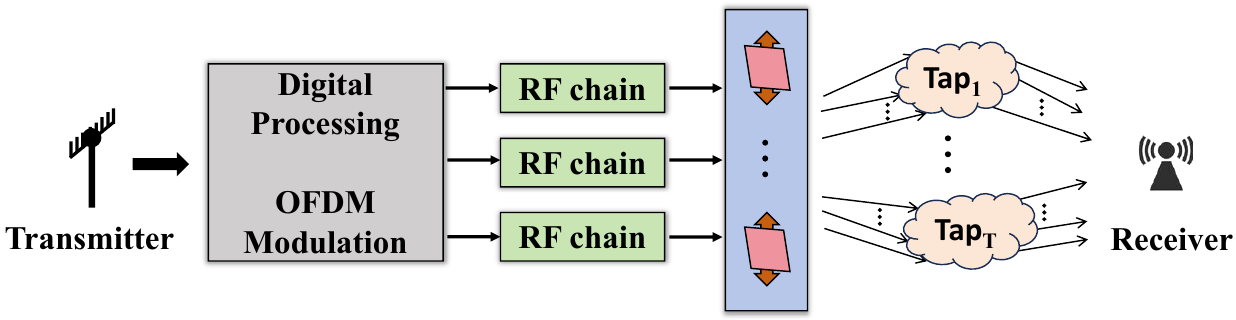} 
  \caption{MA-enhanced MISO OFDM communication system.}
  \label{fig:model}
  \vspace{-6pt}
\end{figure}
However, the aforementioned studies mainly concentrate on the narrow-band MA systems under frequency-flat channels. In the general wideband systems with frequency-selective channels, MA position optimization turns out to be particularly challenging due to the frequency-flat nature of MA positioning, which needs to accommodate the varying channel conditions across different frequency bands. To investigate the performance of MAs in wideband communications, the authors in \cite{zhu2024performance} proposed a parallel greedy ascent (PGA) algorithm to obtain a suboptimal MA position in a SISO system. Moreover, the authors in \cite{cao2025channel} combined the orthogonal matching pursuit algorithm with the least-squares method to estimate CSI across all subcarriers in a wideband MA system. However, there has been no existing works focusing on MA position optimization in wideband MISO systems, to the best of our knowledge. Particularly, the coupling of transmit beamforming and MA position optimization renders the resulting problem even more difficult to be optimally solved compared to the SISO setup considered in \cite{zhu2024performance}. 

To fill in this gap, in this paper, we investigate the MA position optimization in a MISO orthogonal frequency-division multiplexing (OFDM) system, as shown in Fig.\ref{fig:model}, employing the discrete optimization framework similar to \cite{mei2024graph,ma2025robust,wumultidis,mei2024secure,wei2024,weitcom}.In particular, with the optimal maximum-ratio transmission (MRT) in each subcarrier, we aim to jointly optimize the MA positions and power allocation for multiple subcarriers to maximize the achievable rate at the receiver. Although this is a combinatorial problem, we develop an efficient partial enumeration algorithm to find its optimal solution within a branch-and-bound (BB) framework, where a graph-theoretic method is incorporated to effectively prune suboptimal solutions. Moreover, in the low signal-to-noise ratio (SNR) regime, a simplified graph-based algorithm is proposed to obtain the optimal MA positions without the need for enumeration. Numerical results demonstrate that the proposed algorithm outperforms conventional FPAs, while narrowband-based MA position optimization achieves performance close to the optimal.

{\it Notations:} Bold symbols in capital letter and small letter denote matrices and vectors, respectively. The conjugate transpose of a vector or matrix is denoted as ${(\cdot)}^{H}$. ${\mathbb{R}}^n$ (${\mathbb{C}}^n$) denotes the set of real (complex) vectors of length $n$. $\lVert \bf a \rVert$ denotes the Euclidean norm of the vector $\bf a$. For a complex number $s$, $s \sim \mathcal{CN}(0,\sigma^2)$ means that it is a circularly symmetric complex Gaussian (CSCG) random variable with zero mean and variance $\sigma^2$. ${n \choose k} = \frac{n!}{k!(n-k)!}$ denotes the number of ways to choose $k$ elements from a set of $n$ elements. ${\cal O}(\cdot)$ denotes the order of complexity. ${\bf{1}}$ denotes an all-one vector.

\begingroup
\allowdisplaybreaks

\section{System Model and Problem Formulation}
\subsection{System Model}
As shown in Fig.\,\ref{fig:model}, we consider an MA-enhanced MISO-OFDM system, where the transmitter is equipped with $N_t$ MAs and the receiver is equipped with a single fixed antenna. All transmit MAs can flexibly alter their positions within a linear array of length $A$ in meters (m). To facilitate practical deployment, the linear array is uniformly sampled into $M$ discrete positions, with an equal spacing of $\delta = A / (M-1)$. Let $\mathcal{M} = \{0, \delta, 2\delta, \dots, A\}$ denote the set of all sampling positions. The position of the $n$-th MA is denoted by $a_n \in \mathcal{M}, n \in \mathcal{N} = \{1, \dots, N_t\}$. To avoid mutual coupling among MAs, we consider that any pair of transmit antennas must satisfy a minimum separation constraint:
\begin{equation}
|a_i - a_j| \geq a_{\min}, \quad \forall i \ne j,\; i,j \in \mathcal{N},
\end{equation}
where $a_{\min}$ is the minimum allowable distance between any pair of MAs.

We assume that the system operates over a total bandwidth of $B$ Hertz (Hz), which is divided into $L$ OFDM subcarriers. A cyclic prefix (CP) of length $M_{\text{CP}}$ is appended to each OFDM symbol to combat inter-symbol interference caused by multipath propagation. Note that to accommodate the maximum channel delay spread, the cyclic prefix length should satisfy $M_{\text{CP}} \geq T$, where $T$ denotes the number of resolvable channel taps. Let $c_l(a_n)$ denote the channel frequency response (CFR) between the $n$-th transmit MA and the receiver on subcarrier $l$, with $l \in {\cal L} \triangleq \{1, 2, \dots, L\}$. By stacking $c_l(a_n)$ over all $L$ subcarriers, we can obtain the CFR between the $n$-th transmit MA and the receiver over all subcarriers as
\begin{equation}
\mathbf{c}(a_n) = [c_1(a_n), c_2(a_n), \dots, c_L(a_n)]^\top \in \mathbb{C}^{L \times 1}.
\end{equation}
Moreover, by stacking $c_l(a_n)$ over all MAs, the MISO channel on subcarrier $l$ is given by
\begin{equation}
\mathbf{c}_l(\{a_n\}) = \big[c_l(a_1), c_l(a_2), \dots, c_l(a_{N_t})\big]^\top \in \mathbb{C}^{N_t \times 1}.
\end{equation}
To characterize the performance limit of the considered wideband MA system, we assume that all required CSI is perfectly known by applying the existing channel estimation approaches dedicated to MAs (see e.g., \cite{cao2025channel}).

\subsection{Problem Formulation}
Our objective is to maximize the average achievable rate at the receiver over the $L$ subcarriers by jointly optimizing the MA positions and power allocation over the $L$ subcarriers. Let $\mathbf{w}_l \in \mathbb{C}^{N_t}$ denote the transmit beamforming on subcarrier $l$. The associated optimization problem is formulated as
\begin{subequations} \label{eq:P1}
\begin{align}
\text{(P1)} \;
\max_{\{a_n\}, \{\mathbf{w}_l\}} \; 
& \frac{1}{L + M_{\text{CP}}} \sum_{l=1}^{L} 
  \log_2\left(1 + \frac{|\mathbf{w}_l^H \mathbf{c}_l(\{a_n\})|^2}{\sigma^2} \right) \notag \\
\text{s.t.} \; 
& \sum_{l=1}^L \|\mathbf{w}_l\|_2^2 \leq P_{\text{max}}, \label{eq:P1_power} \\
& a_n \in \mathcal{M}, \quad \forall n \in \mathcal{N}, \label{eq:P1_discrete} \\
& |a_i - a_j| \geq a_{\min}, \; \forall i \ne j,\; i,j \in \mathcal{N}, \label{eq:P1_spacing}
\end{align}
\end{subequations}
where $P_{\max}$ denotes the maximum transmit power.

Let $p_l$ denote the transmit power on subcarrier $l$, i.e., $\|\mathbf{w}_l\|_2^2=p_l, l \in {\cal L}$. Then, for any given MA positions $\{a_n\}$ and power allocations $\{p_l\}$, the optimal beamforming vector per subcarrier can be obtained as the maximum-ratio transmission (MRT), i.e.,
\begin{equation}
\mathbf{w}_l = \sqrt{p_l} \cdot \frac{\mathbf{c}_l(\{a_n\})}{\|\mathbf{c}_l(\{a_n\})\|_2}, \quad \forall l \in {\cal L}.
\label{eq:mrt}
\end{equation}

By substituting \eqref{eq:mrt} into (P1), it reduces to the following joint antenna position and power allocation problem,
\begin{subequations} \label{eq:P2}
\begin{align}
\text{(P2)} \quad
\max_{\{a_n\}, \{p_l\}} \; 
& \frac{1}{L + M_{\text{CP}}} \sum_{l=1}^{L} 
  \log_2\left(1 + \frac{p_l \left\| \mathbf{c}_l(\{a_n\}) \right\|_2^2}{\sigma^2} \right) \notag\\
\text{s.t.} \quad 
& \sum_{l=1}^L p_l \leq P_{\text{max}}, \label{eq:P2_power} \\
& p_l \geq 0, \quad \forall l \in \{1,\dots,L\}, \label{eq:P2_nonnegative} \\
& \text{\eqref{eq:P1_discrete}, \eqref{eq:P1_spacing}} \nonumber
\end{align}
\end{subequations}

However, problem (P2) remains a challenging mixed-integer non-convex problem. In the following section, we propose an optimal solution to (P2) via a BB framework with proper solution pruning.\vspace{-3pt}

\section{Proposed Algorithm for (P2)}
To optimally solve (P2), we adopt a BB framework, which enumerates feasible MA positions for (P2) with proper solution pruning. In particular, we employ a graph-based algorithm to determine an upper bound on the optimal value of (P2) for solution pruning. The details are presented as follows. \vspace{-3pt}

\subsection{Feasible Solution Enumeration}
To efficiently enumerate all feasible solutions to (P2), we construct a directed acyclic graph (DAG), denoted by \(G = (V, E)\), where \(V = \mathcal{M}\) is the set of all candidate positions, and \(E\) is the set of edges. In particular, for any two vertices $i,j \in V$, we add an edge from $i$ to $j$ if $j-i \ge a_{\min}$. Hence, the edge set is expressed as
\begin{equation}
E=\{(i,j)|\, i,j \in V,\, j-i \ge a_{\min}\}.
\label{eq:edge_constraint}
\end{equation}

It follows that any feasible solution to (P2) corresponds to an $N_t$-vertex path in $G$. Hence, we can enumerate all feasible solutions to (P2) in a recursive manner. Specifically, let \(\Omega_r, \ r \le N_t\) denote the set of all paths of length \(r\) in \(G\). Obviously, we have \(\Omega_1 = V\). Then, for any path in \(\Omega_r\), denoted as \(\Gamma_r = [\bar{a}_1,\dots,\bar{a}_r]\), we can append any vertex in the following set to obtain a path of length \(r+1\), i.e.,
\begin{equation}
{\cal N}(\bar{a}_r)=\{i|\,i\in V, (\bar{a}_r,i) \in E\},
\end{equation}
which is the set of the outgoing neighbors of vertex $\bar{a}_r$. Note that we set ${\cal N}(\emptyset)=V$. As such, we obtain a new path of length $r+1$ and add it to the set $\Omega_{r+1}$. If ${\cal N}(\bar{a}_r)=\emptyset$, the expansion of $\Gamma_r$ will be terminated. 

By this means, we can construct the path set $\Omega_{N_t}$, in which each element corresponds to a feasible solution to (P2). For each path or antenna position solution in $\Omega_{N_t}$, e.g., $\Gamma_{N_t} = [\bar{a}_1,\dots,\bar{a}_{N_t}]$, it can be shown that the optimal power allocation can be obtained via water-filling as
\begin{equation}
p_l(\Gamma_{N_t}) = \left(\mu - \frac{\sigma^2}{\|\mathbf{c}_l(\{\bar{a}_n\})\|_2^2}\right)^+, \quad \forall l,
\label{eq:waterfilling}
\end{equation}
where \(\mu\) is a water level that can be determined based on \(\sum_{l=1}^L p_l(\Gamma_{N_t}) = P_{\max}\) via bisection search. Then, the achievable rate by $\Gamma_{N_t}$ is given by
\begin{equation}
R(\Gamma_{N_t}) \!= \! \sum_{l=1}^L \log_2\left(\left(\frac{\mu\|\mathbf{c}_l(\{\bar{a}_n\})\|_2^2}{\sigma^2}\right)^+\right),
\label{eq:rate_full}
\end{equation}
where we have omitted the constant scalar $\frac{1}{L+M_{\mathrm{CP}}}$.

Based on \eqref{eq:rate_full}, we can compare the achievable rates by different paths in $\Omega_{N_t}$ and obtain the optimal antenna position solution as
\begin{equation}
\{a_n^\star\} = \arg\max\limits_{\Gamma \in \Omega_{N_t}}R(\Gamma).
\label{eq:opt_path}
\end{equation}

However, the above recursive enumeration results in the complexity in the order of $M-(a_{\min}-1)(N_t-1) \choose N_t$, which may not be applicable to a large size of antenna array with large $M$ and/or $N_t$ values in practice. To avoid this high complexity, we employ a solution bounding method to prune the expansion of certain sub-paths in $\Omega_r, r < N_t$, as presented next.

\subsection{Feasible Solution Bounding}
For any sub-path $\Gamma_r = [\bar{a}_1,\dots,\bar{a}_r]$, a rate upper bound is computed by a frequency-wise relaxation. Specifically, we independently compute the maximum end-to-end channel power gain on each subcarrier under $a_i=\bar{a}_i, 1 \le i \le r$. In this case, the channel power gain on subcarrier $l$ can be expressed in terms of the remaining $N_t-r$ MA positions, i.e., $\{a_n\}_{n=r+1}^{N_t}$, as
\begin{equation}
\gamma_l(\{a_n\}_{n=r+1}^{N_t};\{\bar{a}_i\}_{i=1}^r) 
= \sum_{i=1}^{r} |c_l(\bar{a}_i)|^2 
+ \sum_{n=r+1}^{N_t} |c_l(a_n)|^2.
\label{eq:gamma_actual}
\end{equation}

Based on \eqref{eq:gamma_actual}, its maximum value with respect to (w.r.t.) $\{a_n\}_{n=r+1}^{N_t}$ can be obtained by solving the following optimization problem,
\begin{align}
\max_{\{a_n\}_{n=r+1}^{N_t}} \quad 
& \sum_{n=1}^r \lvert c_l(\bar{a}_n)\rvert^2 + \sum_{n=r+1}^{N_t} \lvert c_l(a_{n})\rvert^2 \label{eq:upperbound_problem}\\
\text{s.t.} \; 
& (\bar{a}_r,a_{r+1}) \in E,\notag\\
&(a_t,a_{t+1}) \in E,\; r+1 \le t \le N_t-1.\notag
\end{align}
Note that the first term in \eqref{eq:upperbound_problem} is a constant. For \eqref{eq:upperbound_problem}, we can obtain its optimal solution in polynomial time by applying a similar graph-based algorithm as in our previous work\cite{mei2024graph}. Thus, we only outline the main steps of solving it. Specifically, we add a ``dummy'' vertex $M+1$ to $G$ and add an edge from each vertex in $V$ to vertex $M+1$. As such, the vertex set becomes $\tilde V = V \cup \{M+1\}$, while the edge set $E$ becomes
\begin{equation}
\tilde E=E \cup \{(j,M+1)|j \in V\}.
\end{equation}
Moreover, for the new graph $\tilde G=(\tilde V,\tilde E)$, we set the weights of its edge $(i,j), (i,j) \in \tilde E$ as $\tilde W_{i,j}=-\lvert c_l(a_i) \rvert^2$. It is not difficult to see that problem \eqref{eq:upperbound_problem} is equivalent to finding the shortest $(N_t-r)$-hop path from vertex ${\bar a}_r$ to vertex $M+1$ in $\tilde G$. This fixed-hop shortest path can be found by applying dynamic programming, leading to a complexity order of ${\cal O}((N_t-r)M^2)$\cite{cheng2004finding}.

By solving \eqref{eq:upperbound_problem} for all $L$ subcarriers, we can obtain the per-subcarrier maximum channel power gains under $a_i=\bar{a}_i, 1 \le i \le r$, denoted as $\gamma_{l,\max}(\{\bar{a}_i\}_{i=1}^r), l \in \cal L$. Hence, we can obtain an achievable rate upper bound for the sub-path $\Gamma_r$ as
\begin{align}
\bar{R}_{\mathrm{ub}}(\Gamma_r) &= \sum_{l=1}^L \log_2\left(1+\frac{p_l \gamma_{l,\max}(\{\bar{a}_i\}_{i=1}^r)}{\sigma^2}\right)\notag\\
&\le  \log_2\left(\left(\frac{\mu_0\gamma_{l,\max}(\{\bar{a}_i\}_{i=1}^r)}{\sigma^2}\right)^+\right)\notag\\
&\triangleq R_{\mathrm{ub}}(\Gamma_r)
\label{eq:R_ub}
\end{align}
where $R_{\mathrm{ub}}(\{\bar{a}_i\}_{i=1}^r)$ in \eqref{eq:R_ub} is obtained by applying water-filling with $\mu_0$ denoting the water level. 

As such, we can employ $R_{\mathrm{ub}}(\{\bar{a}_i\}_{i=1}^r)$ as an upper bound on the achievable rate for any sub-path $\Gamma_r$ in $\Omega_r$. Let $R_{\max}$ denote the currently maximum achievable rate by the paths in $\Omega_{N_t}$. Then, if $R_{\mathrm{ub}}(\{\bar{a}_i\}_{i=1}^r) \le R_{\max}$, we can terminate the expansion from the sub-path $\Gamma_r$. Notably, a larger $R_{\max}$ would help reduce the enumeration complexity of the proposed algorithm. Hence, we can determine $R_{\max}$ as the achievable rate by a low-complexity heuristic antenna position optimization algorithm (e.g., that presented in Section \ref{lowSNR}). In Algorithm 1, we summarize the main procedures of our proposed BB algorithm with graph-based performance bounding.
\begin{algorithm}
\caption{BB Algorithm for Solving (P2)}
\begin{algorithmic}[1]
\State Initialize $R_{\max}$.
\State $\Gamma_0 \gets \emptyset$, $r \gets 0$.
\State Execute BB($\Gamma_0$).
\Procedure{BB}{$\Gamma_r$}
    \If{${\cal N}(\bar{a}_r)=\emptyset$}
        \State \textbf{return}
    \EndIf
    \If{$r < N_t$}
        \State Compute $R_{\mathrm{ub}}(\Gamma_r)$ based on \eqref{eq:R_ub}.
        \If{$R_r^{\mathrm{ub}}(\Gamma_r) \le R_{\max}$}
            \State \textbf{return}
        \EndIf
        \For{each $\bar{a}_{r+1}$ in ${\cal N}(\bar{a}_r)$}
            \State Execute BB($\Gamma_r \cup \bar{a}_{r+1}$).
        \EndFor
    \Else
        \State Compute $R(\Gamma_{N_t})$ based on \eqref{eq:rate_full}.
        \If{$R(\Gamma_{N_t}) > R_{\max}$}
            \State Update $R_{\max} \gets R(\Gamma_{N_t})$
            \State Update $\Gamma_{\mathrm{opt}} \gets \Gamma_{N_t}$
        \EndIf
    \EndIf
\EndProcedure
\State \Return $\Gamma_{\mathrm{opt}}, R_{\max}$
\end{algorithmic}
\end{algorithm}
\subsection{Special Case in Low-SNR Regime}\label{lowSNR}
In this section, we consider a special case with a low SNR, in which the optimal solution to (P2) can be obtained in polynomial time. Specifically, in the low-SNR regime, we have
\begin{equation}
\frac{p_l \left\| \mathbf{c}_l(\{a_n\}) \right\|_2^2}{\sigma^2} \rightarrow 0.
\end{equation}

By employing the following approximation for logarithmic functions, i.e.,
\begin{equation}
\log_2(1 + x) \approx \frac{x}{\ln 2}, \; x \to 0,
\label{eq:low_snr_approx}
\end{equation}
we can obtain an approximate achievable rate as
\begin{equation}
R_{\text{approx}}(\{a_n\}) = \frac{1}{\ln 2} \sum_{l=1}^{L} \frac{p_l \left\| \mathbf{c}_l(\{a_n\}) \right\|_2^2}{\sigma^2}.\label{approx}
\end{equation}

It follows from \eqref{approx} that in the low-SNR regime, for any given MA positions $\{a_n\}$, all transmit power should be allocated to the subcarrier with the maximum channel power gain, i.e., $l^*=\arg\max\nolimits_{l \in \cal L}\left\| \mathbf{c}_l(\{a_n\}) \right\|_2^2$. Hence, (P2) can be simplified as
\begin{subequations} \label{eq:P2}
\begin{align}
\text{(P2-low-SNR)} \quad
\max_{\{a_n\}} \; \max_{l \in \cal L}\;
& \frac{P_{\max} \left\| \mathbf{c}_l(\{a_n\}) \right\|_2^2}{\sigma^2} \\
\text{s.t.} \; \text{\eqref{eq:P1_discrete}, \eqref{eq:P1_spacing}}, \nonumber
\end{align}
\end{subequations}
where the constant ${1}/{\ln 2}$ is omitted. Note that (P2-low-SNR) is similar to the antenna position optimization in a narrowband MA system (for subcarrier $i^*$). Nonetheless, increasing the number of subcarriers helps increase the optimal value of (P2-low-SNR) thanks to the improved frequency diversity.

To solve (P2-low-SNR), we first maximize the channel power gain for each subcarrier $l, l \in \cal L$, assuming $l^*=l$. Denote by $\{a_{n,l}\}$ and $\gamma_{l,\max}=\lVert \mathbf{c}_l(\{a_{n,l}\}) \rVert_2^2$ the optimal antenna position solution and the resulting maximum channel power gain for subcarrier $l$, both of which can be efficiently obtained by invoking the graph-based algorithm proposed in \cite{mei2024graph}. As such, the optimal value of (P2-low-SNR) is given by $\gamma_{l_0,\max}$, with $l_0=\arg\max\nolimits_{l \in \cal L} \lVert \mathbf{c}_l(\{a_{n,l}\}) \rVert_2^2$. The optimal antenna positions for (P2-low-SNR) are given by $\{a_{n,l_0}\}$.

\section{Numerical Results}
This section presents numerical results to demonstrate the performance of the proposed algorithm for MA-enhanced MISO-OFDM systems. The wideband MISO channel is modeled using the field-response channel model presented in \cite{zhu2024performance}. Specifically, the time-domain channel consists of \(T\) resolvable delay taps, each representing a distinct propagation delay. For each tap, the channel comprises \(L_t\) spatial propagation paths, with their angles of departure (AoDs) independently generated based on the uniform distribution between 0 and $2\pi$. The total channel power gain per tap is modeled as a CSCG random variable, assumed to be equally distributed among all channel paths within the tap, and decays exponentially w.r.t. the tap delay\cite{zhu2024performance}. The channel impulse response is obtained by stacking all taps and zero-padding them to length \(L\). The frequency-domain channel \(\mathbf{c}_l(\{a_n\})\) on each subcarrier \(l\) is derived by applying an \(L\)-point discrete Fourier transform (DFT) to the padded time-domain response. 

Unless otherwise stated, the simulation parameters are set as follows. We set \(T = 5\) delay taps, each with \(L_t = 10\) paths. The system operates with $L = 64$ OFDM subcarriers and a cyclic prefix of length $M_{\text{CP}} = 6$. The total transmit power is set to $P_{\text{max}} = 46$ dBm, and the noise power on each subcarrier is set to $\sigma^2 = -60$ dBm. The distance between the transmitter and receiver is $d = 10$ m, and the large-scale fading follows a distance-dependent model with a path-loss exponent of $2.2$ and an exponential tap delay decay factor of $2$. The length of the linear transmit array is set to $A = 6\lambda$, where $\lambda$ denotes the wavelength. The center carrier frequency is set as $f_c = 2.4$, leading to $\lambda = 0.125$ m. The linear array is discretized into $M = 36$ candidate positions, and the minimum spacing between any two MAs is $a_{\min} = 0.5\lambda$. The transmitter is equipped with $N_t = 4$ MAs. All simulation results are averaged over 120 independent Monte Carlo trials.

Moreover, for performance comparison, we introduce the following two benchmarks:
\begin{enumerate}
	\item \textbf{Narrowband-Based Design}: In this benchmark, the MA positions are optimized based on the channels at the central frequency $f_c$, by applying the graph-based algorithm proposed in \cite{mei2024graph}. Then, the power allocations are determined via water-filling.
	\item \textbf{FPAs}: In this benchmark, $N$ antennas are symmetrically deployed at the center of the antenna array with an equal spacing of $a{_{\min}}$. The power allocations are determined via water-filling.
\end{enumerate}

Fig.~\ref{fig:sampling_vs_rate} shows the achievable rate versus the number of candidate positions $M$ under different schemes. It is observed that the achievable rate by the proposed algorithm improves with increasing $M$, owing to the finer spatial sampling resolution that enables more precise antenna positioning. However, the performance gain gradually saturates for $M \ge 48$, indicating that a moderate sampling resolution is sufficient to achieve near-optimal performance of continuous MA position optimization. Nonetheless, the optimal rate performance of MAs is observed to be only slightly better than that of the narrowband-based design. This is likely due to the significant power decay w.r.t. the channel tap, which diminishes the variation in channel conditions across different subcarriers.
\begin{figure}[t]
    \centering
    \includegraphics[width=0.85\columnwidth]{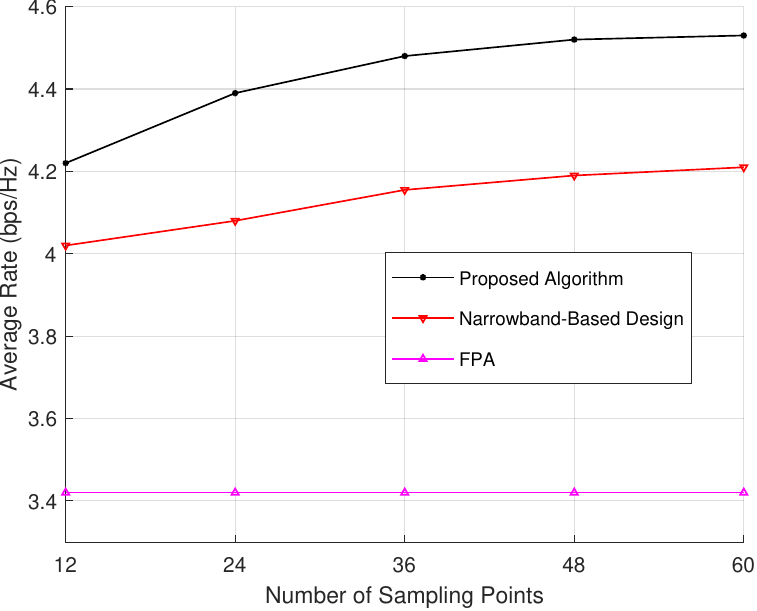}
    \caption{Average rate versus the number of sampling points.}
    \label{fig:sampling_vs_rate}
    \vspace{-6pt}
\end{figure}

Next, we plot the average achievable rate versus the number of transmit antennas $N_t$ in Fig.~\ref{fig:sampling_vs_Nt}. It is observed that the rate performance improves with increasing $N_t$ for all considered schemes, as expected, due to the enhanced beamforming gain provided by more antennas. It is also observed that the performance gap between the proposed algorithm and the FPA benchmark diminishes with increasing $N_t$, as this reduces the flexibility of antenna position optimization given a finite size of the movement region. Moreover, the performance gain of the proposed scheme over the narrowband-based design is observed to remain less than 0.5 bps/Hz. Hence, optimizing the MA positions based on a narrowband channel may be sufficient without incurring significant performance loss.
\begin{figure}[t]
    \centering
    \includegraphics[width=0.85\columnwidth]{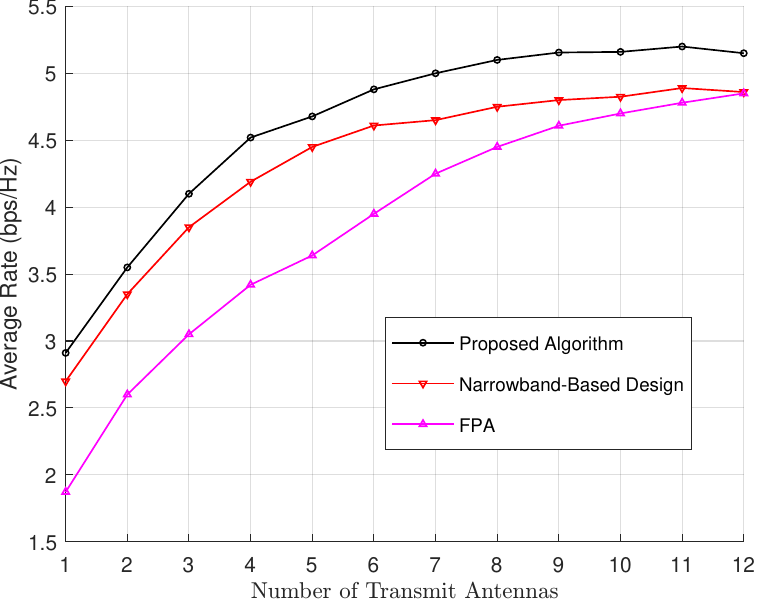}
    \caption{Average rate versus the number of transmit antennas.}
    \label{fig:sampling_vs_Nt}
    \vspace{-6pt}
\end{figure}

Finally, Fig.~\ref{fig:rate_vs_Lt} shows the average rate versus the number of channel paths per tap (i.e., $L_t$). As $L_t$ increases, the rate performance of all considered schemes (except FPAs) is observed to improve, thanks to the enhanced spatial diversity within the antenna array. This thus helps enhance the performance gain of the proposed scheme over the FPA benchmark, as seen from Fig.~\ref{fig:rate_vs_Lt}. For example, the performance gap between them increases from 0.2 bps/Hz to 1 bps/Hz as $L_t$ increases from 1 to 6. Furthermore, the performance gap between the proposed scheme and the narrowband-based design is small throughout the whole range of $L_t$ considered.
\begin{figure}[t]
    \centering
    \includegraphics[width=0.85\columnwidth]{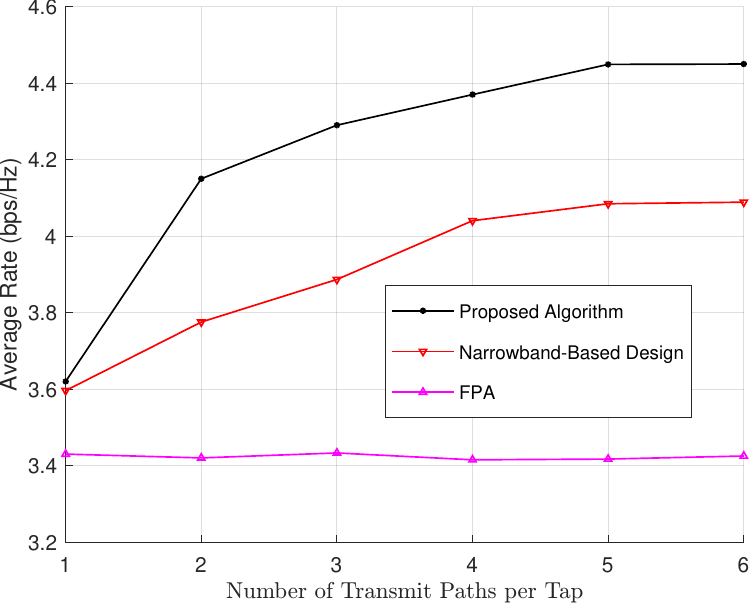}
    \caption{Average rate versus the number of transmit paths per tap.}
    \label{fig:rate_vs_Lt}
    \vspace{-6pt}
\end{figure}

\section{Conclusion}
This paper investigated an MA position optimization problem for a wideband MISO-OFDM system. By discretizing the movement region into a set of candidate positions, we proposed a partial enumeration algorithm to obtain the optimal MA positions within a BB framework. A graph-based method was employed to efficiently eliminate suboptimal solutions in the BB process, and was also used to find the optimal MA positions in the low-SNR regime without requiring enumeration. Simulation results demonstrated that the proposed method achieves a better rate performance than conventional FPAs. However, its performance advantage over the narrowband-based design is small, primarily due to a large power decay across channel taps.

\bibliographystyle{IEEEtran}
\bibliography{references}

\end{document}